# 3D Prestack Fourier Mixed-Domain (FMD) depth migration for VTI media with large lateral contrasts

Zhao, H.*, Gelius, L.-J.*, Tygel, M.**, Harris Nilsen, E.†, and Kjelsrud Evensen, A. †

* University of Oslo, Department of Geosciences, Sem Sælands vei 1, 0371 Oslo, Norway

** University of Campinas, Center for Petroleum Studies, Rua Cora Coralina,

350 Campinus/SP, Brazil

† Lundin Norway AS, Oslo, Norway

Running head: **3D Anisotropic Mixed-domain PSDM**

ABSTRACT

Although many 3D One-Way Wave-equation Migration (OWEM) methods exist for VTI media, most of them struggle either with the stability, the anisotropic noise or the computational cost. In this paper we present a new method based on a mixed space- and wavenumber-propagator that overcome these issues very effectively as demonstrated by the examples. The pioneering methods of phase-shift (PS) and Stolt migration in the frequency-wavenumber domain designed for laterally homogeneous media have been followed by several extensions for laterally inhomogeneous media. Referred many times to as phase-screen or generalized phase-screen methods, such extensions include as main examples of the Split-step Fourier (SSF) and the phase-shift plus interpolation (PSPI). To further refine such phase-screen techniques, we introduce a higher-order extension to SSF valid for a 3D VTI medium with large lateral contrasts in vertical velocity and anisotropy parameters. The method is denoted Fourier Mixed-Domain (FMD) prestack depth migration and can be regarded as a stable explicit algorithm. The FMD technique was tested using the 3D SEG/EAGE salt model and the 2D anisotropic Hess model with good results. Finally, FMD was applied with success to a 3D field data set from the Barents Sea including anisotropy.

# INTRODUCTION

In seismic processing and imaging, the terminology seismic migration refers to methods designed to correct the distortions in position and shape of reflections and diffraction events, in such a way that the transformed (migrated) data are amenable to geological interpretation. Because of its prominent role in extracting meaningful information from seismic data, migration has always been a topic of active research, leading to a large variety of methodologies and applications. A general overview of migration methods, in particular their advantages and disadvantages in theory and practice, can be found in Gray et al. (2001) (see also references therein). Gray and collaborators provide a rough classification of the migration techniques into four main categories: Kirchhoff migration (performed, e.g., by stacking along diffraction curves), finite-difference migration (employing one-way wavefield continuation in space-time or space-frequency domain), reverse-time migration (using finite-differences to solve the full wave equation) and frequency-wavenumber migration (using one-way wavefield continuation in the frequency-wavenumber domain).

As an extension to the latter category, we can define the class of phase-screen propagators that represent a hybrid frequency-wavenumber formulation where also parts of the operations are carried out in the space-domain (a typical example being the thin-lens term). Well-known isotropic algorithms include the Split-Step Fourier (SSF) method (Stoffa et al., 1990) and the Phase Shift Plus Interpolation technique (PSPI) (Gazdag and Sguazzero, 1984). However, SSF degrades severely in accuracy for large velocity contrasts in combination with non-vertically travelling waves. The PSPI can handle lateral velocity variations by using multiple reference velocities within each depth level. However, the accuracy of the method relies on the number of multiple reference functions employed, which again in combination with necessary interpolations unavoidably increases the computational cost. More importantly, an extension of the PSPI method to the anisotropic case represents a major challenge with respect to the construction of an optimized range of reference functions for the anisotropic parameter set. The authors are not aware of any such successful implementation reported. Based on these observations, a new 3D phase-screen propagator scheme is derived in this paper which can handle large contrasts in the anisotropy parameters and the vertical velocity (both laterally and in depth in a VTI medium). The method is denoted Fourier Mixed-Domain (FMD) prestack depth migration (PSDM), due to its combined use of both wavenumber- and space-domain calculations. FMD can be regarded as a stable explicit formulation implemented as a phase-screen operator. For completeness, it should be noted that the higher-order correction terms could alternatively be implemented using an implicit finite-difference scheme. This approach is known in the literature as the Fourier Finite-Difference (FFD) method (Ristow and Rühl, 1994). However, by avoiding a finite-difference implementation in 3D as in the FMD proposed here, numerical anisotropy will not be an issue to cope with (Collino and Joy, 1995). Note also that the extension of the FFD technique to the VTI case is more challenging than that of the implicit FD technique due to difficulties in selecting appropriate references of anisotropy parameters (Hua et al., 2006; Shan, 2009). As pointed out by Zhang and Yao (2012), the choice of the reference anisotropy parameters as the minimum of each layer, will imply construction of a large table of coefficients, whereas the zero-value reference choice will lead to a simplified table but significant loss of accuracy.

The potential superiority of Reverse-Time Migration (RTM) to One-Way Wave-equation Migration (OWEM) in case of complex media is well known. However, RTM is still a costly and computer-intensive technique which typically will be employed in the late stage of processing data from complex models. OWEM is therefore still used as a pragmatic and effective wave-equation depth migration approach and is widely used among the contracting companies for 2D and 3D fast-track depth migration. Thus, the proposed 3D FMD-PSDM technique introduced here should represent a good alternative to current OWEM techniques due to its accuracy and computational efficiency. Moreover, in the velocity model building of complex media, 3D prestack Kirchhoff depth migration is still the 'working horse', due to its computational attractiveness. FMD will outperform Kirchhoff migration in image quality in case of complex geology, and may also be used as an alternative in the iterative velocity-model building due to its computational attractiveness.

This paper is organized as follows. In the first section, we derive the FMD one-way VTI propagator and then introduce the full PSDM scheme. In the section to follow, we discuss the stable implementation of the algorithm in case of strong contrasts in velocity and anisotropy parameters. The FMD method is then tested on controlled data employing the 3D SEG/EAGE salt model and the 2D anisotropic Hess model. In addition, FMD is applied with success to a 3D field data set from the Barents Sea, including anisotropy where the high-velocity target zone representing Permian carbonate rocks is well imaged.

3D FOURIER MIXED-DOMAIN (FMD) ONE-WAY PROPAGATOR FOR A VTI-MEDIUM

With some abuse of notation, we introduce the Fourier transform pairs for a general 3D seismic pressure field $p(\boldsymbol{x},z,t)$ with $\boldsymbol{x}=(x,y)$ representing a position vector in the horizontal plane

$$p(\boldsymbol{x},z,\omega)=\Im_t\left\{p(\boldsymbol{x},z,t)\right\}=\int_{-\infty}^{\infty}dt\exp(-i\omega t)\,p(\boldsymbol{x},z,t),$$
$$p(\boldsymbol{x},z,t)=\Im_\omega^{-1}\left\{p(\boldsymbol{x},z,\omega)\right\}=\left(\frac{1}{2\pi}\right)\int_{-\infty}^{\infty}d\omega\exp(i\omega t)\,p(\boldsymbol{x},z,\omega), \quad (1)$$

and

$$p(\boldsymbol{k},z,\omega)=\Im_x\left\{p(\boldsymbol{x},z,\omega)\right\}=\int_{-\infty}^{\infty}\int_{-\infty}^{\infty}d\boldsymbol{x}\exp[-i\boldsymbol{k}\cdot\boldsymbol{x}]\,p(\boldsymbol{x},z,\omega),$$
$$p(\boldsymbol{x},z,\omega)=\Im_{\boldsymbol{k}}^{-1}\left\{p(\boldsymbol{k},z,\omega)\right\}=\left(\frac{1}{2\pi}\right)^2\int_{-\infty}^{\infty}\int_{-\infty}^{\infty}d\boldsymbol{k}\exp[i\boldsymbol{k}\cdot\boldsymbol{x}]\,p(\boldsymbol{k},z,\omega), \quad (2)$$

with $\boldsymbol{k}=(k_x,k_y)$ representing the wavenumber vector. Our aim is to back-propagate $p(\boldsymbol{x},z,\omega)$ from level $z_j$ to $z_{j+1}=z_j+\Delta z$ by downward extrapolation in the frequency and dual space-wavenumber domains. In symbols, we assume that $p(\boldsymbol{x},z_j,\omega)$ is known and wish to find an approximation of $p(\boldsymbol{x},z_{j+1},\omega)$.

The starting point is the following ansatz for a mixed-domain representation of the vertical wavenumber (dispersion relation)

$$k_{zj}(\boldsymbol{x},\boldsymbol{k},\omega)=\sqrt{\frac{k_j^2(\boldsymbol{x})-(1+2\varepsilon_j(\boldsymbol{x}))k_T^2}{1-2\left[\varepsilon_j(\boldsymbol{x})-\delta_j(\boldsymbol{x})\right]k_T^2/k_j^2(\boldsymbol{x})}} \quad (3)$$

with

$$k_j = \frac{\omega}{c_j(\boldsymbol{x})} \quad , \quad \boldsymbol{k}\cdot\boldsymbol{k} = \sqrt{k_x^2 + k_y^2} = k_T^2 \tag{4}$$

Note that in equation 3, the positive sign in front of the square root corresponds to the backpropagation (migration) case. Correspondingly, forward propagation is obtained by introducing a negative square root. In equation 4, ω is a fixed angular frequency, $\varepsilon_j(\boldsymbol{x})$ and $\delta_j(\boldsymbol{x})$ are the Thompson parameters , and $c_j(\boldsymbol{x})$ is the laterally varying vertical medium velocity within the j-th layer. We assume that evanescent waves are removed in equation 3, namely that $k_j^2(\boldsymbol{x}) - [1 + 2\varepsilon_j(\boldsymbol{x})]k_T^2 \geq 0$.

In case of no anisotropy, equation 3 takes the form of the mixed-domain representation as proposed by Margrave (1998) and Margrave and Ferguson (1999) for the isotropic case within the framework of nonstationary filter theory. In case of a constant medium, equation 3 resembles the dispersion relation introduced by Alkhalifah (1998) for a VTI medium.

Let $\boldsymbol{x'}$ and $\boldsymbol{x}$ represent position vectors in the horizontal plane at input level $z_j$ and output level $z_j + \Delta z$, respectively. Based on equation 3, the following one-way wavefield extrapolation scheme can be constructed:

$$p(\boldsymbol{x}, z_j + \Delta z, \omega) = \Im_{\boldsymbol{k}}^{-1}\left[\Im_{\boldsymbol{x'}}\left\{p(\boldsymbol{x'}, z_j, \omega)\right\}\cdot \exp\left[ik_{Z_j}(\boldsymbol{x}, \boldsymbol{k}, \omega)\Delta z\right]\right] . \tag{5}$$

Equation 5 can be regarded as a generalization of the continuous-velocity PSPI algorithm of Ferguson and Margrave (2002) to the anisotropic case. The name continuous-velocity PSPI is given with reference to the original PSPI-method of Gazdag and Sguazzero (1984).

To achieve efficient implementation of the algorithm in equation 5, we seek to factorize the dispersion relation in equation 3 in separate wavenumber and spatial terms. We begin by introducing the globally optimized cascaded form of the VTI dispersion relation to second order (Zhang and Yao, 2012)):

$$k_{zj}(\boldsymbol{x},\boldsymbol{k},\omega) \cong k_j(\boldsymbol{x})\left[1+\xi-\frac{a_j(\boldsymbol{x})k_T^2/k_j^2(\boldsymbol{x})}{1-b_j(\boldsymbol{x})k_T^2/k_j^2(\boldsymbol{x})}\right] \tag{6}$$

with coefficients defined as

$$\xi = -0.00099915 \quad , \quad a_j(\boldsymbol{x}) = 0.46258453(1+2\delta_j(\boldsymbol{x})) \quad ,$$
$$b_j(\boldsymbol{x}) = 2(\varepsilon_j(\boldsymbol{x})-\delta_j(\boldsymbol{x})) + 0.40961897(1+2\delta_j(\boldsymbol{x})) \tag{7}$$

We are seeking a solution to equation 5 which allows a split into a background plane-wave term associated with a layered model and additional correction terms taking into account lateral velocity variations and higher dip angles. This approach is by analogy with the well-known Split-Step Fourier (SSF) method of Stoffa et al. (1990). Thus, we introduce a constant background or reference medium characterized by the parameters $\{c_{0j}, \varepsilon_{0j}, \delta_{0j}\}$ and with a corresponding dispersion relation:

$$k_{z0j}(\boldsymbol{k},\omega)=\sqrt{\frac{k_{0j}^2-(1+2\varepsilon_{0j})k_T^2}{1-2\left[\varepsilon_{0j}-\delta_{0j}\right]k_T^2/k_{0j}^2}}\cong k_{0j}\left[1+\xi-\frac{a_{0j}k_T^2/k_{0j}^2}{1-b_{0j}k_T^2/k_{0j}^2}\right]\quad,\quad k_{0j}=\frac{\omega}{c_{0j}} \tag{8}$$

where

$$a_{0j}=0.462584531\left(1+2\delta_{0j}\right),\ b_{0j}=2(\varepsilon_{0j}-\delta_{0j})+0.40961897(1+2\delta_{0j}). \tag{9}$$

We also introduce the following useful relation:

$$k_j(\boldsymbol{x})=\sqrt{1+\gamma_j(\boldsymbol{x})}\cdot k_{0j}\quad,\quad \gamma_j(\boldsymbol{x})=\frac{c_{0j}^2}{c_j^2(\boldsymbol{x})}-1 \tag{10}$$

with $\gamma_j$ being the scattering potential or velocity contrast. By the use of equation 10 and a Taylor expansion (finite number of terms N assumed), we can approximate equation 3 as follows:

$$k_{zj}(\boldsymbol{x},\boldsymbol{k},\omega)\cong k_j(\boldsymbol{x})\left[1+\xi-\frac{a_j(\boldsymbol{x})k_T^2/k_j^2(\boldsymbol{x})}{1-b_j(\boldsymbol{x})k_T^2/k_j^2(\boldsymbol{x})}\right]=k_{0j}\sqrt{1+\gamma_j(\boldsymbol{x})}\cdot \left[1+\xi-\frac{A_j(\boldsymbol{x})k_T^2/k_{0j}^2}{1-B_j(\boldsymbol{x})k_T^2/k_{0j}^2}\right]=k_{0j}\sqrt{1+\gamma_j(\boldsymbol{x})}\cdot \left[1+\xi-\frac{A_j(\boldsymbol{x})k_T^2/k_{0j}^2}{\left\{1-b_{0j}k_T^2/k_{0j}^2\right\}\left\{1-\frac{(B_j(\boldsymbol{x})-b_{0j})k_T^2/k_{0j}^2}{\left[1-b_{0j}k_T^2/k_{0j}^2\right]}\right\}}\right]\cong k_{0j}\sqrt{1+\gamma_j(\boldsymbol{x})}\cdot \left[1+\xi-\sum_{n=0}^{N}\frac{A_j(\boldsymbol{x})(B_j(\boldsymbol{x})-b_{0j})^n(k_T^2/k_{0j}^2)^{n+1}}{\left\{1-b_{0j}k_T^2/k_{0j}^2\right\}^{n+1}}\right] \tag{11}$$

where

$$A_j(\boldsymbol{x}) = \frac{a_j(\boldsymbol{x})}{1+\gamma_j(\boldsymbol{x})} \quad , \quad B_j(\boldsymbol{x}) = \frac{b_j(\boldsymbol{x})}{1+\gamma_j(\boldsymbol{x})}$$

(12)

Next, we introduce the equation

$$k_{zj}(\boldsymbol{x},\boldsymbol{k},\omega) \cong k_{z0j}(\boldsymbol{k},\omega) + \left[k_{zj}(\boldsymbol{x},\boldsymbol{k},\omega) - k_{z0j}(\boldsymbol{k},\omega)\right]_{approx}$$

(13)

where the quantities inside the bracket are calculated employing equation 8 and 11 giving as a final result:

$$k_{zj}(\boldsymbol{x},\boldsymbol{k},\omega) \cong k_{z0j}(\boldsymbol{k},\omega) + \left[k_j(\boldsymbol{x}) - k_{0j}\right](1+\xi) +$$

$$k_{0j}\left[\frac{\left\{a_{0j} - \sqrt{1+\gamma_j(\boldsymbol{x})}A_j(\boldsymbol{x})\right\}k_T^2/k_{0j}^2}{\left\{1 - b_{0j}k_T^2/k_{0j}^2\right\}} - \sum_{n=1}^{N}\frac{\sqrt{1+\gamma_j(\boldsymbol{x})}A_j(\boldsymbol{x})\left\{B_j(\boldsymbol{x}) - b_{0j}\right\}^n(k_T^2/k_{0j}^2)^{n+1}}{\left\{1 - b_{0j}k_T^2/k_{0j}^2\right\}^{n+1}}\right]$$

(14)

The three terms on the right-hand side of equation 14 can now be easily identified as the background term, the modified thin-lens term and a higher-order correction term of order N.

To test the robustness of the approximation given by Equation 14, we calculated the relative dispersion error as a function of phase or propagation angle (no lateral variation in parameters). We considered two cases: (i) weak-contrast case with $\frac{c_0}{c} = \frac{\varepsilon_0}{\varepsilon} = \frac{\delta_0}{\delta} = 4/5$ and a (ii) strong-contrast case with $\frac{c_0}{c} = \frac{\varepsilon_0}{\varepsilon} = \frac{\delta_0}{\delta} = 1/2$. In both simulations, we let $\varepsilon = 0.3$ , $\delta = 0.1$, but the velocity c changed from 2,500m/s to 4,000m/s between the two runs. It can be easily seen from Figure 1 that the strong contrast case performs almost as well as the weak-contrast case and that the 1% phase-error limit is around 55–60 degrees. Due to the use of an *optimized* version of the anisotropic dispersion relation to second order, it may happen that for a given combination of anisotropy parameters, the weak-contrast case will locally perform slightly more poorly than the strong-contrast case (e.g. in the current example for the largest angles).

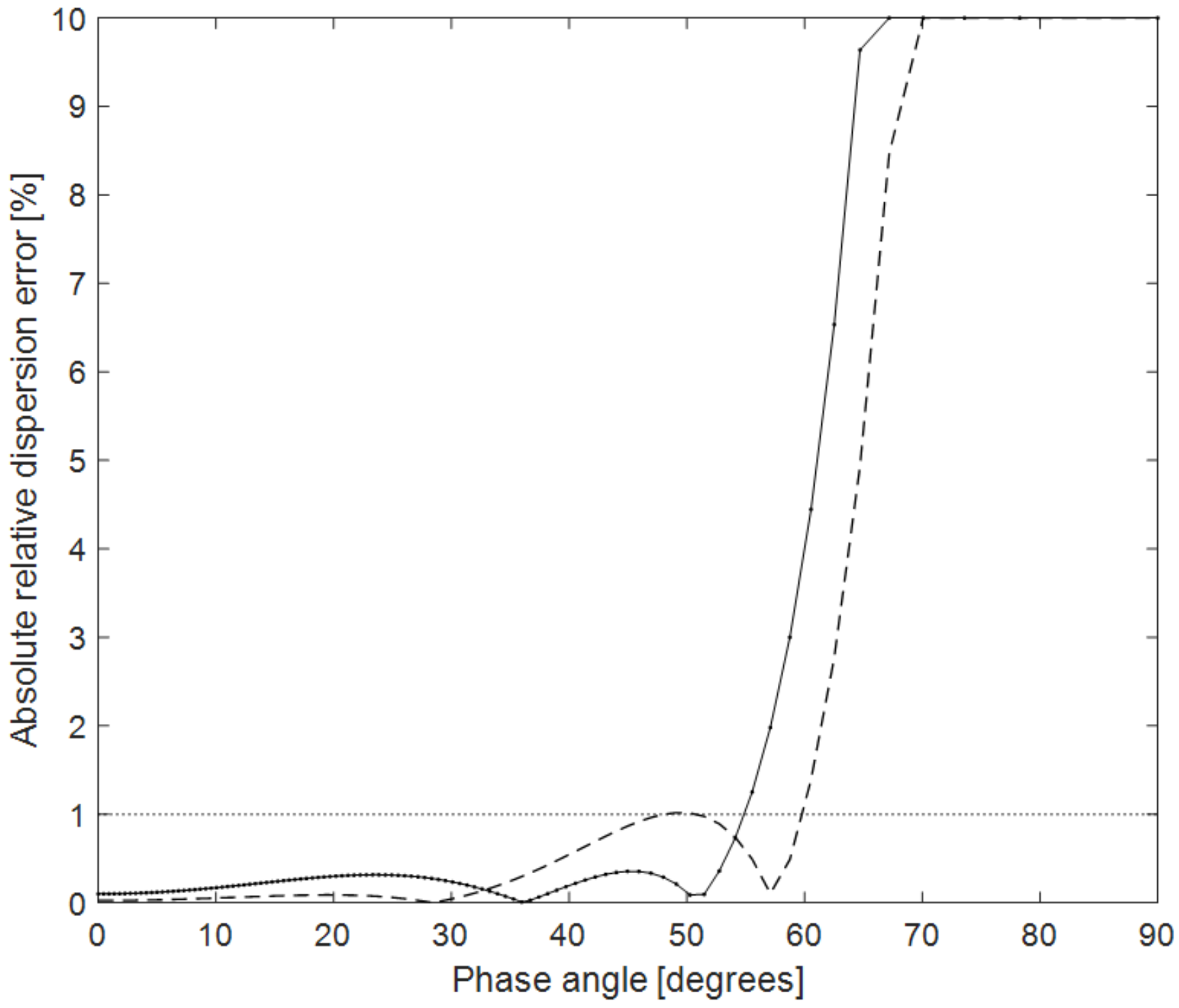

*Figure 1. Relative dispersion error versus phase angle: weak-contrast case (solid line) and strong-contrast case (broken line). The 1 % dispersion-error line has also been superimposed.*

Improved accuracy can be obtained if the analytical expressions for the parameters $A_j$ and $B_j$ are replaced by parameter fitting at higher angles formulated as an optimization problem. Such an approach is used by Shan (2009) to obtain optimized implicit finite-difference schemes for VTI media. However, because our ultimate goal is to carry out 3D prestack PSDM in complex geological models, such an optimization approach will be a highly time-consuming task, building several predefined tables of coefficients. However, such tables are typically built once for a given dataset and applied to all the shots in the survey. In this paper, we also avoid a finite-difference implementation of the perturbation term in equation 14 but introduce a stable explicit propagator in the Fourier mixed-domain. Accordingly, we denote our method as Fourier Mixed-Domain (FMD) PSDM. By avoiding a finite-difference implementation in 3D, numerical anisotropy will not be an issue to cope with (Collino and Joy, 1995) as in Ristow and Rühl (1994). These latter authors derived an alternative expression for the dispersion relation in equation 14 with one perturbation term and implemented this term as a cascading Fourier finite-difference (FFD) operator (implicit and stable scheme).

Based on equation 14, a one-way VTI propagator can now be constructed:

$$exp\left[ik_{zj}(\boldsymbol{x},\boldsymbol{k},\omega)\Delta z\right] = exp\left[ik_{z0j}(\boldsymbol{k},\omega)\Delta z\right] \cdot exp\left[i(k_j(\boldsymbol{x}) - k_{0j})(1+\xi)\Delta z\right] \cdot$$
$$exp\left[\frac{k_T^2/k_{0j}^2}{\left\{1 - b_{0j}k_T^2/k_{0j}^2\right\}}\left\{ik_{0j}\left(a_{0j} - \sqrt{1+\gamma_j(\boldsymbol{x})}A_j(\boldsymbol{x})\right)\Delta z\right\}\right] \cdot$$
$$exp\left[-\sum_{n=1}^{N}\frac{(k_T^2/k_{0j}^2)^{n+1}}{\left\{1 - b_{0j}k_T^2/k_{0j}^2\right\}^{n+1}}\left\{ik_{0j}\sqrt{1+\gamma_j(\boldsymbol{x})}A_j(\boldsymbol{x})\left(B_j(\boldsymbol{x}) - b_{0j}\right)^n\Delta z\right\}\right] \tag{15}$$

The two last exponential factors on the right-hand side of equation 15 are approximated using a first-order Taylor expansion, an approach which leads to the following symbolic version of a mixed-domain VTI PSDM scheme (after reorganization and neglecting high-order cross-terms):

$$P(z_j + \Delta z) = exp\left[ik_{z0j}\Delta z\right] exp\left[i(k_j(\boldsymbol{x}) - k_{0j})(1+\xi)\Delta z\right]\left\{1 + \Gamma_j(u,\boldsymbol{x})\right\}P(z_j) \quad ,$$

$$\Gamma_j(u,\boldsymbol{x}) = \left\{ik_{0j}\left\{a_{0j} - \sqrt{1+\gamma_j(\boldsymbol{x})}A_j(\boldsymbol{x})\right\}\Delta z\right\}\frac{u}{\left[1 - b_{0j}u\right]} -$$

$$\sum_{n=1}^{N} ik_{0j}\sqrt{1+\gamma_j(\boldsymbol{x})}A_j(\boldsymbol{x})\left[B_j(\boldsymbol{x}) - b_{0j}\right]^n \Delta z \frac{u^{n+1}}{\left[1 - b_{0j}u\right]^{n+1}} \quad , \quad u = k_T^2/k_{0j}^2 \tag{16}$$

To make the explicit formulation in equation 16 unconditionally stable, we introduce a dip-filter $\Psi_j(u)$ defined by the condition

$$\Psi_j(u) = \frac{1}{max(\left|1 + \Gamma_j(u,\boldsymbol{x})\right|)} \quad , \quad 0 \le u \le 1 \tag{17}$$

To ensure that this dip-filter harms the data as little as possible, reference values of the vertical velocity and the anisotropy parameters are computed using the mean values. In the explicit migration of Hale (1991), a stability constraint similar to equation 17 is employed but in the space domain. It should be noted that the conventional FD method is not stable when the medium velocity has sharp discontinuities (Biondi, 2002). Zhang et al. (2003) also use similar ideas to stabilize an isotropic phase-screen migration scheme.

The final version of the FMD scheme now takes the form:

$$
\begin{aligned}
p(\boldsymbol{x}, z_j + \Delta z, \omega) \\
&= exp\left[i(k_j(\boldsymbol{x}) - k_{0j})(1+\xi)\Delta z\right] \Im_{\boldsymbol{k}}^{-1}\left\{exp\left[ik_{z0j}\Delta z\right] \Psi_j(k_T^2/k_{0j}^2) \Im_{\boldsymbol{x}'}\left[p(\boldsymbol{x}', z_j, \omega)\right]\right\} \\
&+ \\
\left\{ik_{0j}\left\{a_{0j} - \sqrt{1+\gamma_j(\boldsymbol{x})}A_j(\boldsymbol{x})\right\}\Delta z\right\} exp\left[i(k_j(\boldsymbol{x}) - k_{0j})(1+\xi)\Delta z\right] \cdot \\
\Im_{\boldsymbol{k}}^{-1}\left\{exp\left[ik_{z0j}\Delta z\right] \Psi_j(k_T^2/k_{0j}^2) \frac{k_T^2/k_{0j}^2}{\left[1 - b_{0j}k_T^2/k_{0j}^2\right]} \Im_{\boldsymbol{x}'}\left[p(\boldsymbol{x}', z_j, \omega)\right]\right\} - \\
\sum_{n=1}^{N} ik_{0j}\sqrt{1+\gamma_j(\boldsymbol{x})}A_j(\boldsymbol{x})\left[B_j(\boldsymbol{x}) - b_{0j}\right]^n \Delta z\left[exp\left[i(k_j(\boldsymbol{x}) - k_{0j})(1+\xi)\Delta z\right] \cdot \right. \\
\left. \Im_{\boldsymbol{k}}^{-1}\left\{exp\left[ik_{z0j}\Delta z\right] \Psi_j(k_T^2/k_{0j}^2) \frac{(k_T^2/k_{0j}^2)^{n+1}}{\left[1 - b_{0j}k_T^2/k_{0j}^2\right]^{n+1}} \Im_{\boldsymbol{x}'}\left[p(\boldsymbol{x}', z_j, \omega)\right]\right\}\right]
\end{aligned}
\tag{18}
$$

In case of an isotropic medium, the parameters in equation 18 take the simplified forms

$$
a_{0j} = 0.46258453 \quad , \quad b_{0j} = 0.40961897 \quad , \quad A_j = \frac{a_{0j}}{(1+\gamma_j(\boldsymbol{x}))} \quad , \quad B_j = \frac{b_{0j}}{(1+\gamma_j(\boldsymbol{x}))} \tag{19}
$$

**Dual-reference model**

Our overall goal is to develop a reconstruction (migration) scheme that is able to image complex geological models (e.g., with the inclusion of salt diapers), and at the same time being computationally attractive. Due to its formulation, the FMD technique fulfils the last criterion, but inaccuracies in phases are to be expected in case of very strong vertical-velocity contrasts (i.e., velocity jumps of a factor of three and more) and/or similar large contrasts in the anisotropic parameters. In order to handle such more extreme cases, we propose dual-reference FMD for which the basic idea is as follows:

- if a region exists within a given depth-migration strip where the velocity and/or the anisotropy contrasts are larger than a user-defined factor (e.g. 2.5), backpropagation employing FMD is carried out twice for that extrapolation depth: first with the mean values as the references and second with a parameter set chosen as the mean of the values of the anomalous region(s);
- for such a migration strip, the two results are finally merged at the output level in the space domain.

The above conditions can be mathematically described as

$$p(\boldsymbol{x}, z_j + \Delta z, \omega) = \sum_{i=1}^{2} \mathrm{M}_i \left(\boldsymbol{x}\right) p_i(\boldsymbol{x}, z_j + \Delta z, \omega), \tag{20}$$

where $p_i$ (i = 1,2) represents the extrapolated field using as a reference velocity field the mean-velocity of the non-anomalous regions (say, i=1) and the anomalous regions (i=2) respectively. Moreover, as in the equation, $\mathrm{M}_i$ denote corresponding mask functions as follows: If i=1 specifies the mean velocity, then $\mathrm{M}_1$, as in the equation, takes the value 1 at all location points corresponding to the non-anomalous regions and 0 otherwise. Correspondingly, the second mask-function $\mathrm{M}_2$ represents the complementary case, $\mathrm{M}_2 = 1\text{-}\mathrm{M}_1$.

## Comparison with literature of screen-propagators

The attractive features of simplicity and computational efficiency of frequency-wavenumber techniques have motivated a series of works aiming to generalize the approach to be valid in a more realistic geological setting. The most popular frequency-wavenumber migration schemes are Phase Shift (Gazdag, 1978) and Stolt f-k migration (Stolt, 1978). Although very quick and inexpensive, both techniques have the drawback of being limited to velocity media that varies only with depth. In order to handle lateral velocity variations, Gazdag and Sguazzero (1984) introduced PSPI. It can handle lateral velocity variations by using multiple reference velocities within each depth level. However, the accuracy of the method relies on the number of multiple reference functions employed, an approach which again in combination with necessary interpolation unavoidably increases the computational cost. An extension of the PSPI method to the anisotropic case represents a major challenge regarding how to construct an optimized range of reference functions of the anisotropic parameter set. The authors are not aware of any such successful implementation being reported.

If we consider the limit of vertically travelling waves (e.g., $k_T \to 0$) and an isotropic case, equation 16 will take the form of the Split-Step Fourier (SSF) method introduced by Stoffa et al. (1990). SSF can handle lateral variations and only requires a single reference velocity for each depth level. The SSF operator is unconditionally stable but degrades in accuracy for large velocity contrasts in combination with non-vertically travelling waves. Popovici (1996) extended SSF to the prestack case formulated in the offset-midpoint domain employing the DSR equation. Jin and Wu (1999) extended this latter work to also include higher-order terms. Still, the combination of strong velocities and steep angles is not treated in an accurate manner. Within an isotropic formulation, other higher-order alternatives to the SSF technique have been proposed to cope with larger propagation angles. Huang et al. (1999) introduced the Extended Local Born Fourier (ELBF) propagator to include waves propagating at non-vertical angles and Chen and Ma (2006) proposed a higher-order version of ELBF. However, despite being able to handle larger angles more accurately, this class of screen-propagators still suffers from the underlying Born assumption in case of larger velocity contrasts and the propagators become unstable in use in the frequency-wavenumber domain. Le Rousseau and de Hoop (2001a) introduced an isotropic higher-order scheme which they denoted Generalized Screen (GS) propagators. The GS scheme is more robust to velocity variations than ELBF type of schemes, but all of these techniques suffer from a singularity at the evanescent boundary. Le Rousseau and de Hoop proposed a phase normalization to stabilize the algorithm, but the accuracy of this normalization degrades with the complexity of the model (only exact for a homogeneous model). In an accompanying work, Le Rousseau and de Hoop (2001b) extended the GS scheme to a VTI type of medium. However, only weak approximations of the anisotropy parameters were introduced, and the demonstration part was limited to one modelling example (thus, no imaging results provided).

Note that all techniques discussed above are restricted to a range of propagation angles when it comes to accuracy. Thus, they do not perform better than FMD in terms of this issue in case of an isotropic medium.

**3D common-shot implementation of FMD**

Applied to each common-shot gather, the FMD migration follows the classical procedure (see Claerbout, 1971) of computing, as a first step, the frequency-domain reflectivity function $r(\boldsymbol{x}, z_j, \omega)$ at all levels $z = z_j$, and next applying the imaging condition of inverse Fourier transforming that reflectivity to the time domain and evaluating it at time zero.

We now explain the algorithm to extrapolate the reflectivity $r(\boldsymbol{x}, z_j, \omega)$ at level $z = z_j$ (supposedly already known) to the (unknown) reflectivity $r(\boldsymbol{x}, z_j + \Delta z, \omega)$ at the new level $z = z_j + \Delta z$. Following Claerbout (1971), an estimate of the reflectivity function $r(\boldsymbol{x}, z_j + \Delta z, \omega)$ can be written in the form

$$r(\boldsymbol{x}, z_j + \Delta z, \omega) = \frac{U(\boldsymbol{x}, z_j + \Delta z, \omega)}{D(\boldsymbol{x}, z_j + \Delta z, \omega)} = \frac{U(\boldsymbol{x}, z_j + \Delta z, \omega) D^*(\boldsymbol{x}, z_j + \Delta z, \omega)}{D(\boldsymbol{x}, z_j + \Delta z, \omega) D^*(\boldsymbol{x}, z_j + \Delta z, \omega)}. \tag{21}$$

Here, $U(\boldsymbol{x}, z_j + \Delta z, \omega)$ and $D(\boldsymbol{x}, z_j + \Delta z, \omega)$ are upward and downward pressure wavefields defined as follows. On one hand, $U(\boldsymbol{x}, z_j + \Delta z, \omega)$ represents the backward extrapolation of the recorded common-source surface data to level $z = z_j + \Delta z$. On the other hand, $D(\boldsymbol{x}, z_j + \Delta z, \omega)$ represents the forward extrapolation of the common-source point wavefield from the surface to level $z = z_j + \Delta z$. We assume that $U(\boldsymbol{x}, z_j, \omega)$ and $D(\boldsymbol{x}, z_j, \omega)$ at level $z = z_j$ are already available. Then, the FMD extrapolations to $U(\boldsymbol{x}, z_j + \Delta z, \omega)$ and $D(\boldsymbol{x}, z_j + \Delta z, \omega)$ can symbolically be expressed as

$$U(\boldsymbol{x}, z + \Delta z, \omega) = \ell_{FMD} U(\boldsymbol{x}, z, \omega) \text{ and } D(\boldsymbol{x}, z + \Delta z, \omega) = \ell^*_{FMD} D(\boldsymbol{x}, z, \omega), \tag{22}$$

where $\ell_{FMD}$ represents the backward FMD propagator and $\ell^*_{FMD}$ the corresponding FMD forward propagator. Thus, we assume that our FMD propagator can be well approximated by a pure-phase or plane-wave propagator. In such a case, the forward propagator is readily obtained from its backward propagator counterpart by means of a simple complex conjugation.

Finally, taking the inverse Fourier Transform over frequency and applying it to t =0 and an additionally a summation over number of shot points, provides the sought-for pre-stack FMD depth migration

$$R(\boldsymbol{x}, z_j + \Delta z) \cong \sum_k \frac{\sum_i U_k(\boldsymbol{x}, z_j + \Delta z, \omega_i) D_k^*(\boldsymbol{x}, z_j + \Delta z, \omega_i)}{\left\langle \sum_i D_k(\boldsymbol{x}, z_j + \Delta z, \omega_i) D_k^*(\boldsymbol{x}, z_j + \Delta z, \omega_i) \right\rangle}, \tag{23}$$

the summations running over all available discrete frequencies (i) and shot points (k). Note that equation 23 represents a slightly different deconvolution imaging condition (IC) than the classical version of Claerbout (1971). The summation over frequencies is here carried out separately for the nominator and denominator in equation 23. We found that this approach gave an improved image in case of the data investigated in this paper. The notation < > in equation 23 indicates smoothing with a triangular filter. Before the smoothing was applied, the spatial average value Iav of the illumination function $I(\boldsymbol{x}, z_j + \Delta z) = \sum_i D_k(\boldsymbol{x}, z_j + \Delta z, \omega_i) D_k^*(\boldsymbol{x}, z_j + \Delta z, \omega_i)$ was calculated and the following threshold introduced: if $I(\boldsymbol{x}, z_j + \Delta z) < 0.2 I_{av}$ then replace it with that value. From experience based on the datasets considered in this study, the scale-factor of 0.2 in this threshold equation seemed to be a good choice. However, in general this scale-factor can be user selected and depend on the data being employed.

## CONTROLLED DATA EXAMPLES

In this section, we demonstrate the ability of the proposed FMD technique to provide accurate imaging results in case of complex controlled models. The first example involves the 3D isotropic SEG/EAGE model, whereas the second study is based on the 2D VTI Hess model.

### 3D isotropic SEG/EAGE model

Data were taken from the SEG/EAGE Salt Model Phase C WA (Wide Azimuth) survey (Aminzadeh et al, 1996). For this data set, each shot has eight streamers with a maximum of 68 groups per streamer. The group interval is 40 m, the cable separation is 80 m and the shot interval is 80 m. The sample interval is 8 ms, the recording length is 5 s and the centre frequency of the source pulse is 20 Hz. The survey consists of 26 sail lines separated by 320 m and with 96 shots per line. In order to properly apply the FK-part of our imaging technique, the original 2D receiver layout corresponding to each shot point was interpolated to a finer and regular grid of 20m x 20m. This interpolation was carried out in the frequency domain employing a 2D spline algorithm. It is likely that the use of the more sophisticated 5D-type of interpolation algorithms would have given even better results. However, the authors did not have access to such techniques. A depth increment of 20m was used in the FMD migration scheme. In this example, we employed the dual-velocity concept and a second-order scattering scheme (corresponding to n=2 in Equation 18).

Figure 2 shows a 3D visualization of the final imaging results based on representative slices through the image cube. The overall quality seems to be quite satisfactory, given the complexity of the model and the imperfectness in the data generation and acquisition geometry.

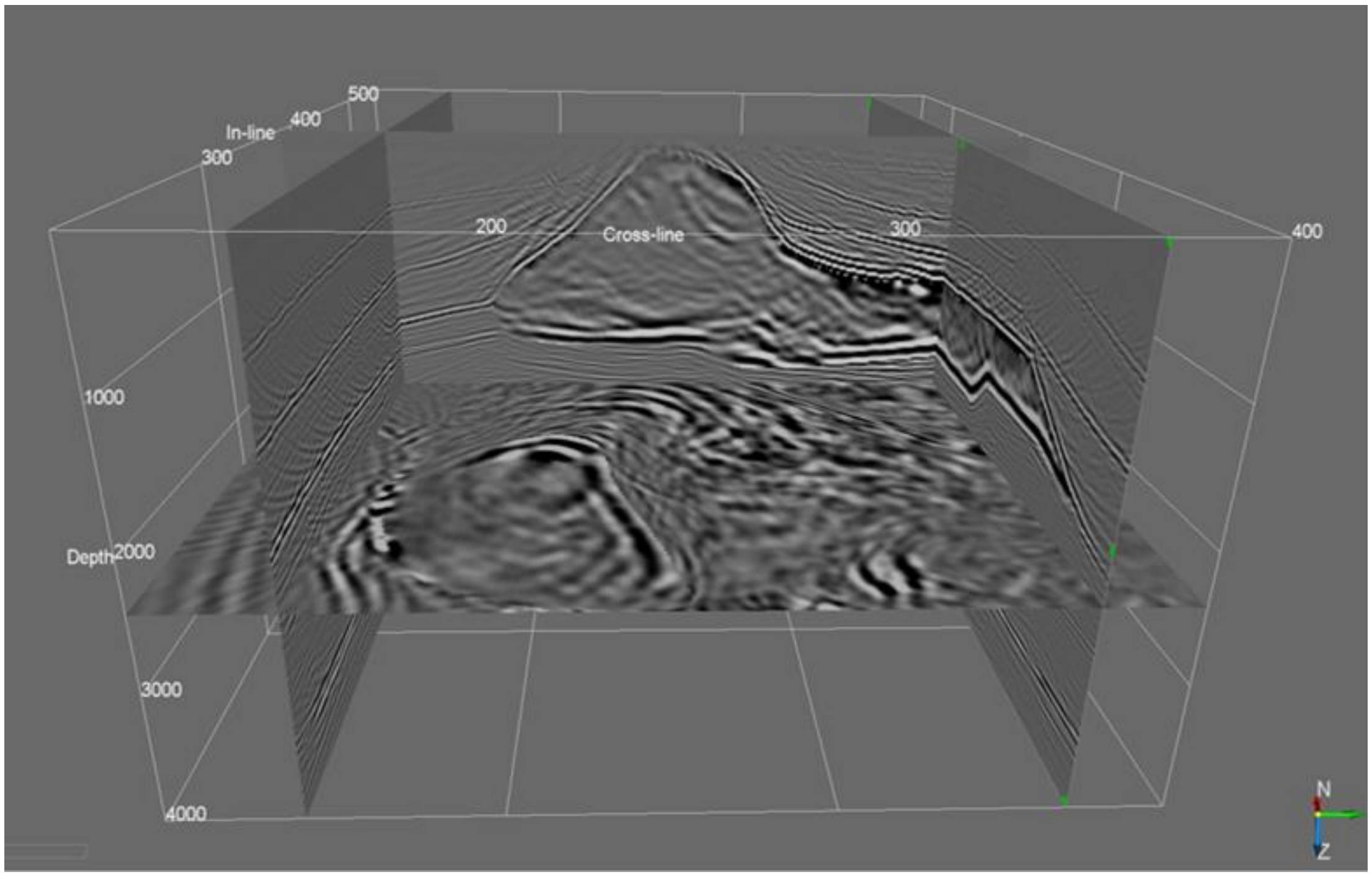

*Figure 2. Slices through the final image cube*

To further address the quality of the FMD technique, we selected the inline profile located at X = 7700m. Figure 3a represents the corresponding velocity model and with the associated imaged line shown in Figure 3b. The obtained reconstruction shown in Figure 3b can now be compared with recently published results in the literature:

- Jang and Kim (2016) gave examples of the implementation under the use of Parallel 3D PSPI migration. The cited work is one of the latest articles of a 3D screen-propagator technique employed to the 3D SEG/EAGE salt model. In this way, it is appropriate to include their results in this paper. Their Figure 6 represents the same inline profile as the one in our Figure 3b. Direct comparison shows that the 3D FMD result is superior with respect to resolution and accuracy. The image obtained by Jang and Kim (2016) demonstrates the difficulties when going from 2D to 3D using Fourier techniques; in particular, the issue of spatial aliasing is a main challenge.

- Li et al. (2015) introduces 3D weak-dispersion RTM using a so-called Stereo-Modeling Operator. They apply the RTM method to the SEG/EAGE Salt Model, and their Figure 5 gives the image of the same profile as in our Figure 3b. However, note that Li et al. (2017) employed data from Phase A, an approach which implies that each shot has six streamers and not eight as in Phase C. More importantly, the coverage of the right part of the model is larger in Phase A. Thus, the most-right part of our image is missing simply because of this lack of coverage. When the relevant parts of the image are compared, our method is superior with more reflectors present.

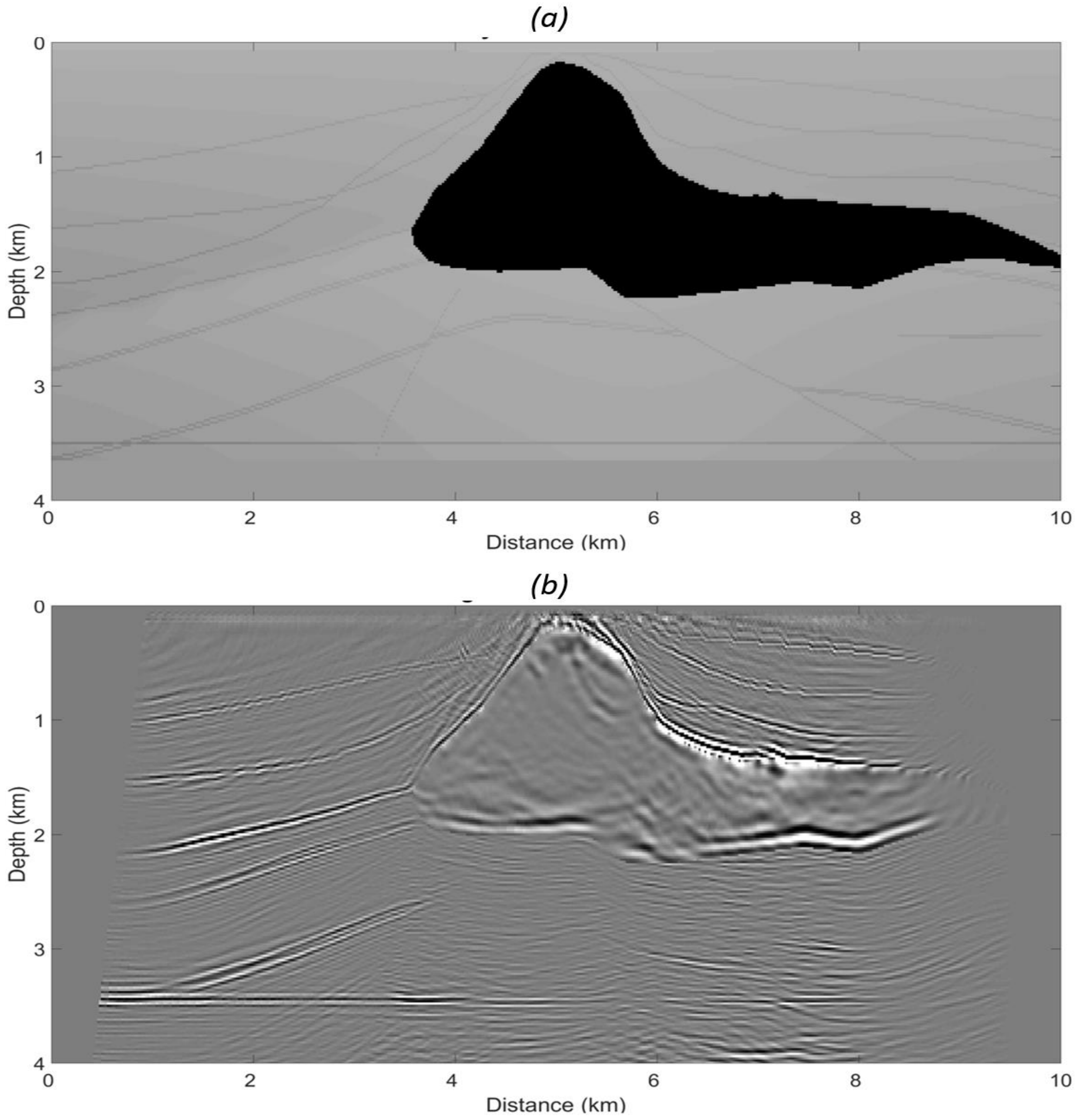

*Figure 3(a) Stratigraphic interval velocity model and (b) corresponding FMD image*

**2D Hess VTI model**

This model was originally built by J. Leveille and F. Qin of Amerada Hess Corp, and is considered to be representative of several exploration areas in the Gulf of Mexico. The overall structural complexity is moderate, but it includes a salt body surrounded by sedimentary layers and a relatively steep fault plane. The magnitudes of the coefficients ε and δ are in some of the intervals considered to fall between moderate to strongly anisotropic. In this study, we employ the multiple-free version of the data generated by SEP at the Stanford University. The data set consists of a 2D line with 720 shots separated by 100 ft and with offsets ranging from 0 ft and 26,200 ft (receiver spacing of 40 ft). The trace length is 8 s and the temporal sampling interval is 6 msec. A depth increment of 20 ft was used in the imaging stage. Figure 4 shows the final image obtained using the VTI-FMD technique. In this example, we used the dual-velocity approach and a first-order scattering approximation (i.e. use of n=1 in Equation 18). The reconstruction is well resolved with respect to both the fault system, the steep salt flank and the anisotropic anomalous regions.

We can also compare the image in Figure 4 with recent results reported in the literature:

- Shin and Byun (2013) implemented the VTI version of the GS scheme (Le Rousseau and de Hoop, 2001a and 2001b) and tested it using the Hess model. Direct comparison with their Figure 7b shows that the FMD technique is superior in quality: better resolved shallow parts and top salt, better-defined faulting system and the ability to image the steep flank of the salt structure. Due to the fact that the VTI-GS scheme is generally regarded as the most optimal one among the phase-screen propagators, the result obtained by our FMD technique is therefore rather encouraging.

- Han et al. (2018) introduced a wavelength-dependent Fresnel-beam migration (FBM) technique valid for VTI media. They applied the wavelength-dependent FBM to the Hess model and obtained the result shown in their Figure 9c. In addition, a comparison with standard FBM was included (cf. their Figure 9b). Han et al. (2018) employed Gaussian smoothing of the model parameters in advance of the migration. Direct comparison with the FMD reconstruction in Figure 4 shows that the two results are very similar in quality, but with FMD recovering more structures at the far-most left part of the model. However, a slight variation in amplitude of events to the right of the salt exists in the FMD result. This is due to a tighter mute applied to the migrated shots closer to the major fault to minimize spurious events.

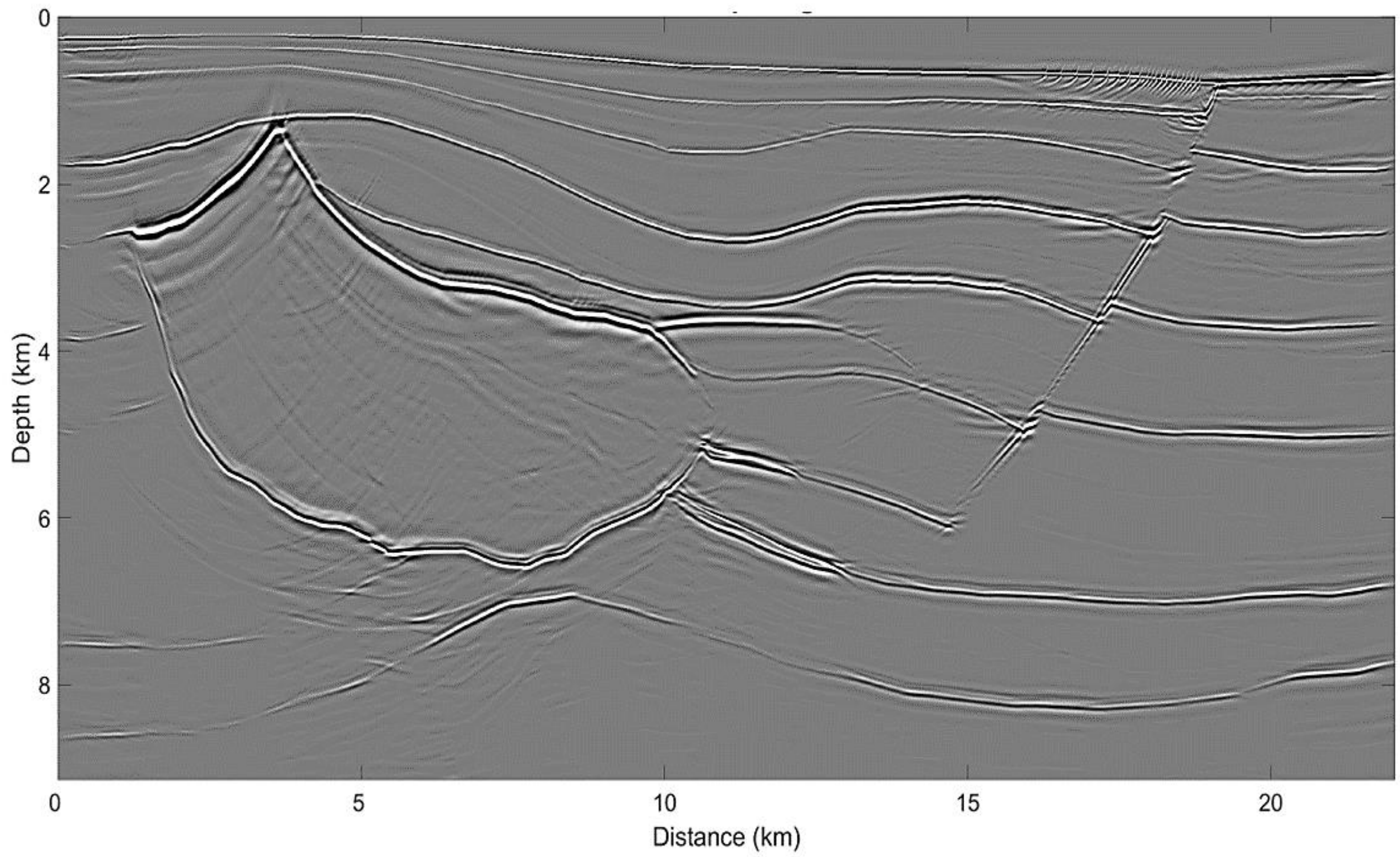

*Figure 4. Image of the Hess model obtained using 2D VTI-FMD.*

## 3D MARINE FIELD DATA EXAMPLE

A 3D marine dataset provided by Lundin Norway AS and acquired in the Barents Sea is used as a benchmark of the full 3D VTI-FMD method. The dataset was acquired with 12 streamers separated by 75 meters and a dual-source configuration. Table 1 provides a summary of the key acquisition parameters.

*Table 1: Key acquisition parameters*

| | | | |
|---|---|---|---|
| **Number of sources** | 2 | **Number of groups** | 564 per streamer |
| **Depth of source arrays** | 6 m | **Streamer separation** | 75 m |
| **Shot point interval** | 18.75 m | **Nominal near offset** | 120.9 m |
| **Number of streamers** | 12 | **Sample rate** | 2 ms |
| **Active streamer length** | 7050 m | **Record length** | 7060 ms |
| **Depth of streamers** | 18.0 – 29.0m +/- 2.0m | **Nominal fold** | 94 |
| **Group interval** | 12.5 m | | |

Several challenges were associated with this field data set. Firstly, strong ocean currents forced the seismic survey to be acquired along the strike direction of the subsurface geology to increase the operational efficiency. However, this approach implied increased challenges for both the 3D seismic processing and imaging due to the increased amount of out-of-plane contributions. The strong ocean current also led to a significant amount of cable feathering. Figure 5 gives an example of cable feathering for one selected shot point, where the feathering is seen to amount to approximately 300 meters or more. Finally, the hard sea floor in the Barents Sea also caused strong noise interference in the marine data set.

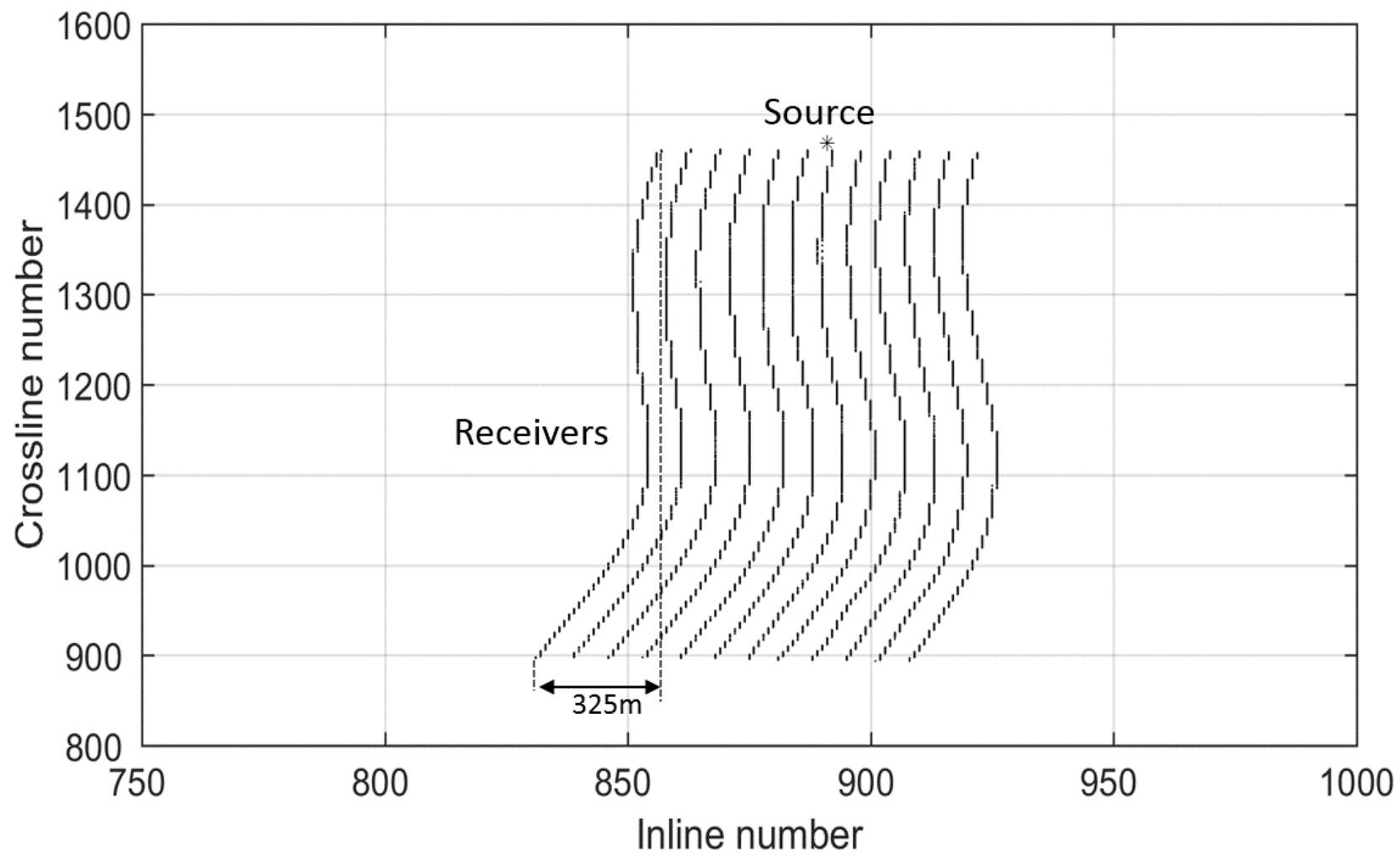

*Figure 5. Example of strong cable feathering for selected shot point.*

The field data had been pre-processed by a contracting company prior to being employed in this study. This pre-processing involved navigation merging, debubble, attenuation of swell noise, and seismic interference noise as well as 3D SRME. The authors, to improve data quality and save computational time, further processed the data set. This additional processing involved the following steps:

- resampling from 2–4 ms,
- application of a tau-p mute to remove residual linear noise,
- bandpass-filtering to keep frequencies between 2–80 Hz only,
- data regularization employing a 2D spline interpolation in the frequency and space domain (12.5 m by 12.5 m inline and crossline sampling after regularization),
- mute in offset keeping only smaller offsets for larger travel times (due to a large increase in velocity from overburden to target zone), and
- keeping only a recording length of 2 s (sufficient to image the main target area).

In case of a real production processing, interpolation using 2D splines should be avoided due to possible smearing effects. Thus, more advanced 5D interpolation algorithms like Minimum Weighted Norm Interpolation (MWNI) (Liu and Sacchi, 2004) or Anti-Leakage Fourier Transform (ALFT) reconstruction (Xu et al., 2005) should be the preferred choice.

Before the actual 3D shot-point migration was executed, an appropriate zero-padding was introduced in space and time to minimize transform and migration noise. The contracting company had provided 3D depth cubes of the vertical velocity, as well as the anisotropy parameters ε and δ (cf. Figure 6). It can be seen from Figure 6 that a significant jump in the vertical velocity characterizes the area at larger depths and that the anisotropy parameters are reflecting the same jump and in general with a simple step-like variation.

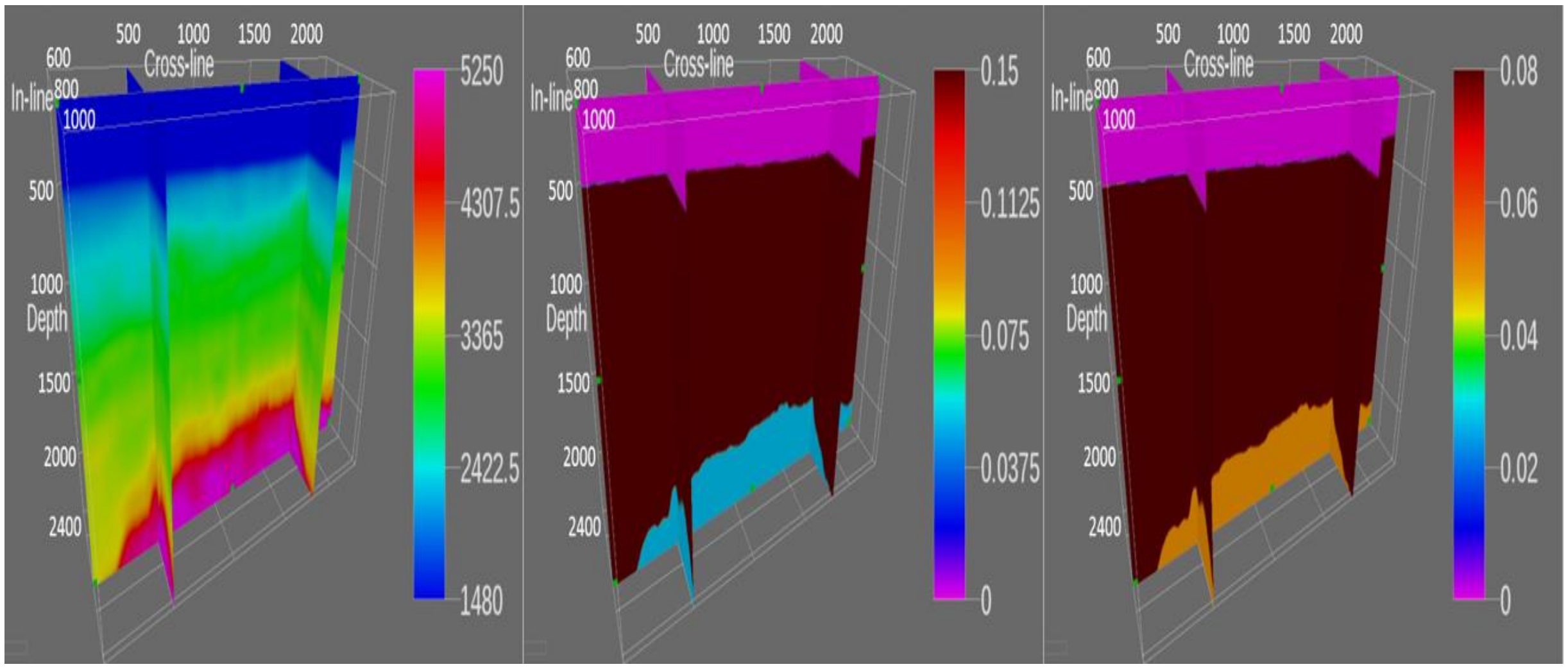

*Figure 6. Vertical-velocity cube, epsilon cube and delta cube (from left to right)*

A depth increment of 4 m was chosen for the migration. In this field data example, we applied a single-velocity approach and a first-order scattering approximation to lower the computational burden. Figure 7 shows a 3D visualization of the final imaging results based on representative slices through the image cube. The fits between the inline and crossline sections seem to be overall good. Note that due to the heavy computational burden associated with 3D VTI type of PSDM using our prototype-software in Matlab, we limited the 3D demonstration of our algorithm to three sail lines in Figure 7. To investigate further the quality of the migrated cube, one representative inline section is shown in Figure 8. This reconstruction is formed by employing a depth-dependent aperture that only included one sail-line in the overburden and a smooth transition to the use of three sail lines within the carbonate target zone. The overall image quality is seen to be highly satisfactory. The overburden is well imaged with its highly-resolved fault systems. The high-velocity target zone starting at Top Permian reconstructs equally well both the top structures and the faulted reflector band below. Because these Permian carbonate rocks represent a major jump in the velocities, only a smaller band of offsets were employed within this zone in order to avoid critically refracted events harming the overall image quality.

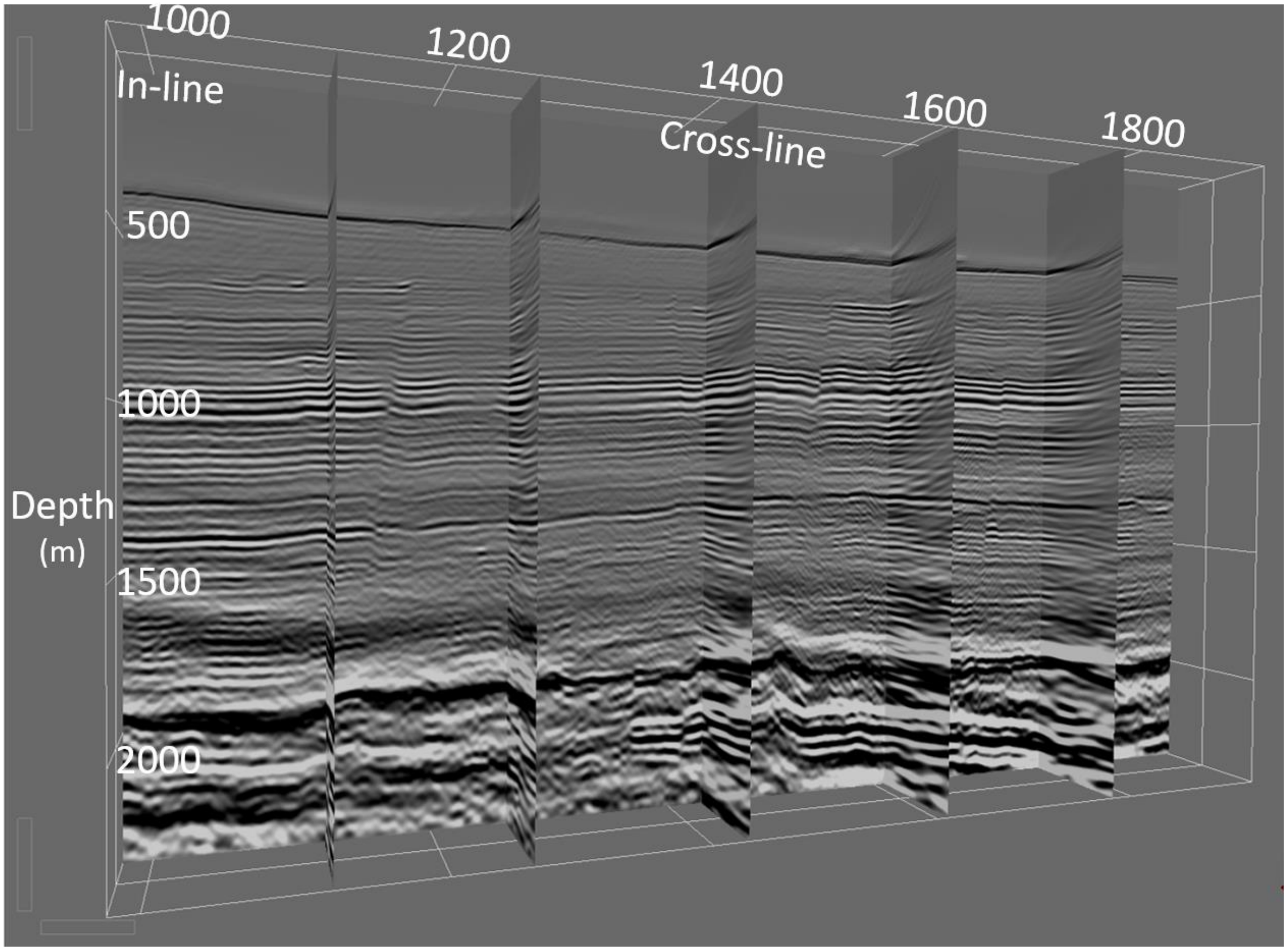

*Figure 7. 3D VTI-FMD depth migration: slices through image cube. Note that the result is based on one sail-line only (cross-line sections stretched for ease of visualization).*

To further demonstrate the good performance of our proposed method, the corresponding image result obtained by the previously mentioned contracting company is shown in Figure 9. The contracting company made use of a sophisticated angle-migration approach implemented in the offset-midpoint domain. Direct comparison between Figures 8 and 9 support our claim regarding the excellent image quality provided by the FMD technique. It represents a better-resolved and less noisy migration except for the left-most part of the image at a depth of approximately 2 km where some dipping noise appears, which is due to the use of a smaller lateral aperture than the contractor.

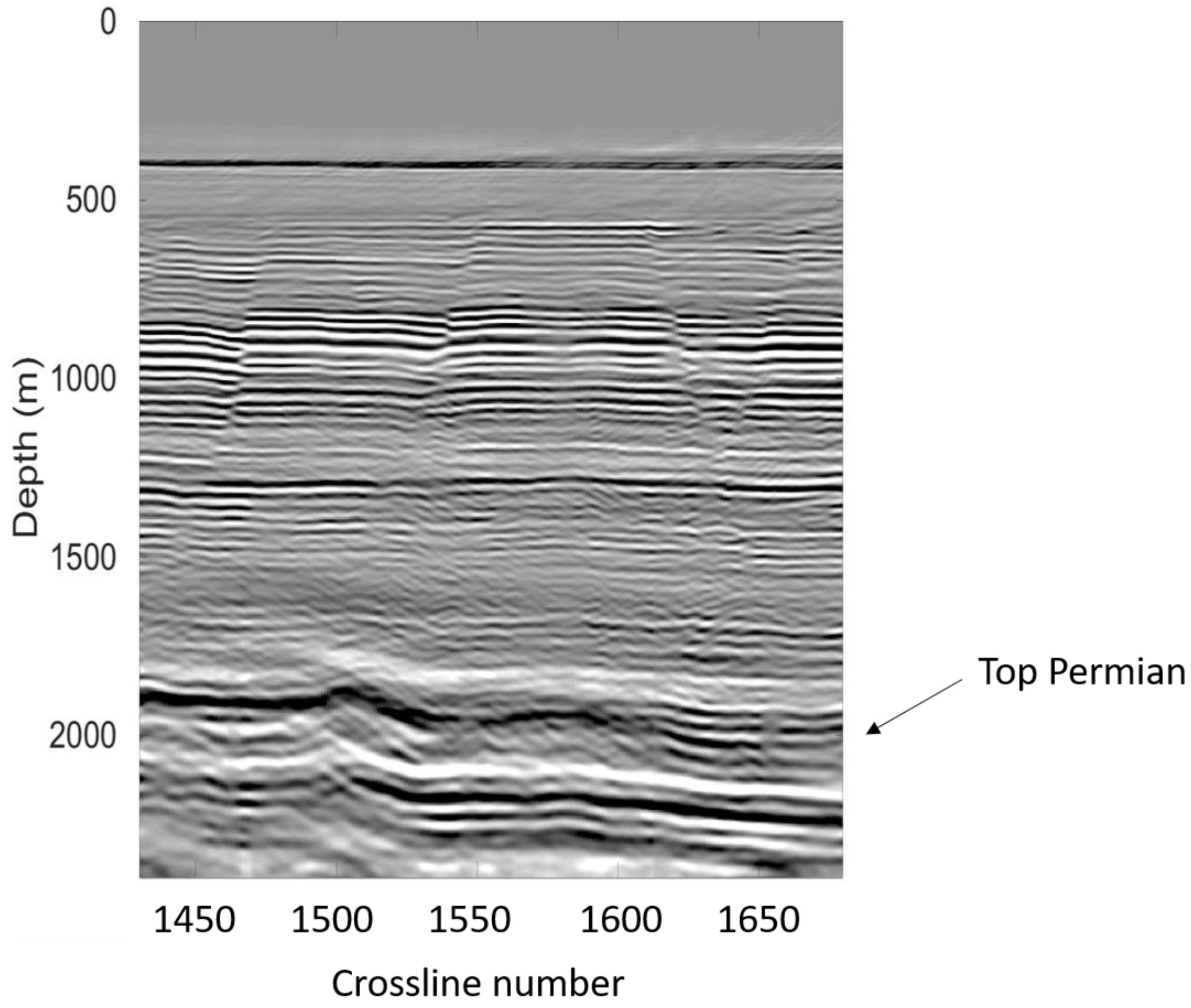

*Figure 8. Inline section 1891 taken from the image cube in Fig.7.*

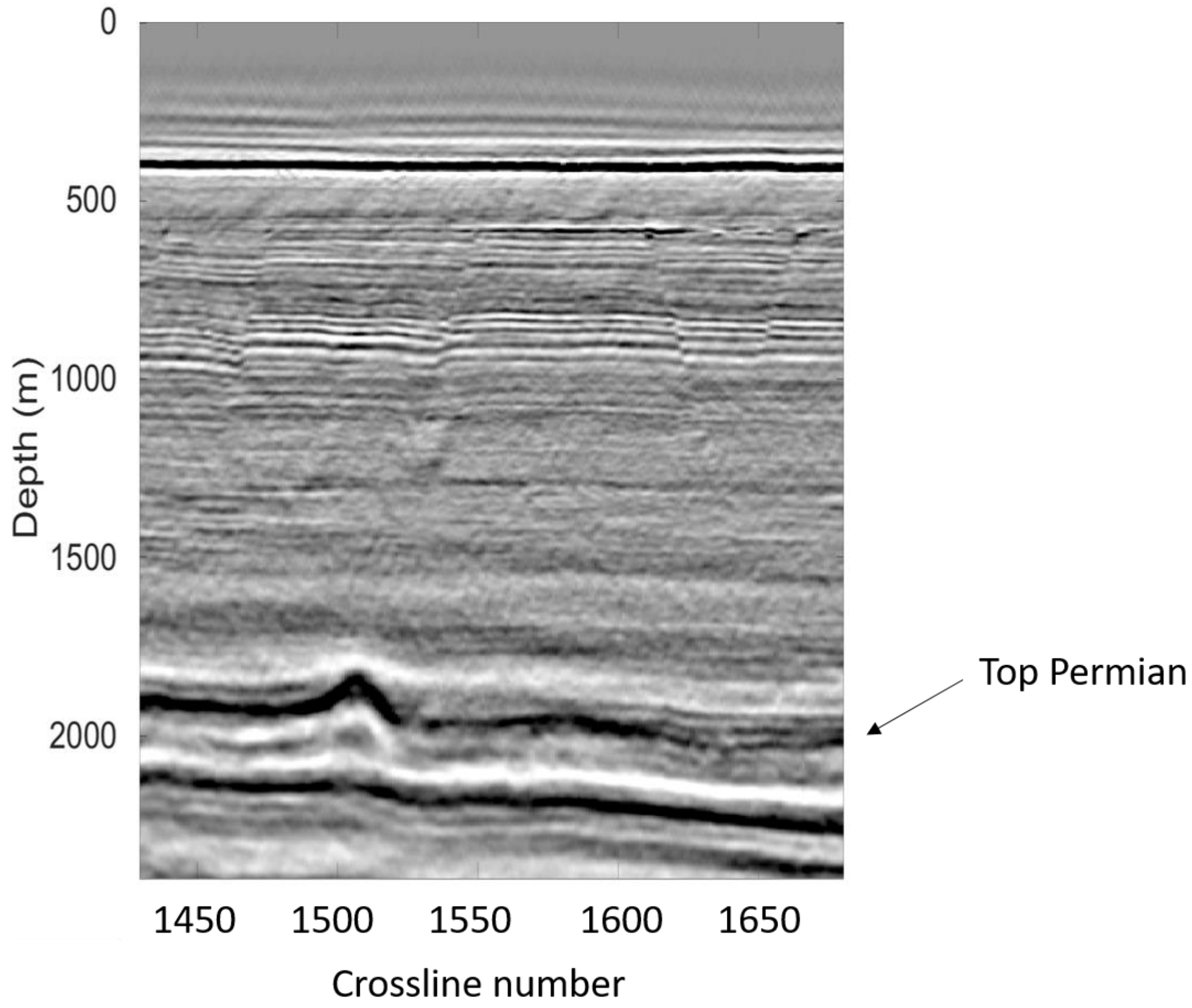

*Figure 9. Same inline section as in Figure 8, but as provided by the contracting company using 3D VTI angle-migration*

## Computational issues

In this work, we have followed common practice and developed a research prototype of our method employing Matlab. To develop a commercial C++ code has been outside the scope of this paper. However, by the use of figures reported from the literature describing typical speed improvements when converting a Matlab code to an optimized C++ version, we can make estimates regarding how well the FMD technique will perform after such a conversion.

The Matlab code ran on a supercomputer consisting of 650+ Supermicro X9DRT computing nodes. All nodes are dual Intel E5-2670 (Sandy Bridge) running at 2.6 GHz, yielding 16 physical cores. Each node has 64 GB of DDR3 memory operating at 1,600 MHz, giving 4 GB memory per physical core at approximately 58 GB/s aggregated bandwidth using all physical cores. Because this super computer is a shared resource for several universities, we only had access to a limited part of its computing capacity (typically not more than 40–60 nodes). The computational time for 100 3D- shots, taken from the field data set, distributed on 20 nodes (16 cores each) was typically about 20 hours, a result which implies that on average, five 3D shots per node consumed the same amount of time.

If a program written in a high-level language such as Matlab is converted to an optimized $C^{++}$ code, we can expect a typical improvement in computational speed in the range of 10–100 based on the experiences reported by professional program developers. Andrews (2012) even reports an improvement in speed of several hundreds. If we employ the conservative factor of 20, it implies that we can compute five 3D shots per node using about 60 minutes (or approximately 12 minutes per 3D shot per node). From a major contracting company, we have been informed that for 3D depth migration based on 1-way formulation, the computational cost for an optimized code with the same source-receiver layout will be typically around eight minutes per 3D shot per node. However, to obtain such a computational speed, the company also *made use of a GPU environment*. Thus, based on this industry example as well as our conservative analysis of possible gain in computational speed within a CPU environment, it is highly likely that a significantly efficient and competitive implementation can be achieved for the FMD method on a $C^{++}$/GPU platform.

## CONCLUSION

In this paper, a new migration technique for 2D and 3D prestack data also valid for vertical transverse isotropic media has been presented. It can be regarded as a higher-order version of the Split-Step Fourier (SSF) method and is denoted Fourier Mixed-Domain (FMD) migration. By applying an optimized dip filter, the FMD is shown to be stable for strong variations in anisotropy and velocity parameters despite being an explicit type of scheme.

In contrast to Fourier Finite Difference migration, the high-order correction terms are implemented as screen-propagator terms, avoiding the issues of anisotropic noise in 3D finite-difference implementations.

The FMD technique was tested using the 3D SEG/EAGE salt model and the 2D anisotropic Hess model with good results. Finally, FMD was applied with success to a 3D field data set from the Barents Sea including anisotropy where the high-velocity target zone representing Permian carbonate rocks was well imaged. Direct comparison with the result obtained by a contracting company using a sophisticated angle-migration technique, further demonstrated the superior image resolution provided by FMD imaging.

The current version of the FMD method can handle 3D VTI media. Further extension to the more general TTI case is ongoing research. In addition to the set of perturbed medium parameters inherent in the present formulation, also the tilt of the symmetry axis needs to be included in a computer efficient manner.

Future potential use of the FMD technique, besides being an efficient prestack depth migration (PSDM) method, could also be in iterative PSDM velocity building as an alternative to the industry-preferred Kirchhoff method.

ACKNOWLEDGEMENTS

H. Z. and L.-J. G. acknowledge support from the Norwegian Research Council through a PETROMAKS 2 project (NFR/234019).

M. T. acknowledges support from the National Council for Scientific and Technological Development (CNPq-Brazil), the National Institute of Science and Technology of Petroleum Geophysics (INCT-GP-Brazil) and the Center for Computational Engineering and Sciences (Fapesp/Cepid # 2013/08293-7-Brazil). He also acknowledges support of the sponsors of the Wave Inversion Technology (WIT) Consortium and the Brazilian Oil Company (Petrobras).

Finally, the authors thank Lundin Norway AS for making the 3D field data set available for this study. The authors also thank SEG and Hess Corporation for providing the SEG and Hess synthetic data.

# REFERENCES

Alkhalifah, T., 1998. Acoustic approximations for processing in transversely isotropic media. Geophysics 63, 623–631.

Aminzadeh, F, Burkhard, N., Long, J., Kunz, T., Duclos, P., 1996. Three dimensional SEG/EAEG models—An update. The Leading Edge 15 (2), 131–134.

Andrews, T., 2012. Computation Time Comparison Between Matlab and C++ Using Launch Windows. American Institute of Aeronautics and Astronautics (6 pages).

Biondi, B., 2002. Stable-wide angle Fourier finite difference downward extrapolation of 3D wavefields. Geophysics 67 (3), 872–882.

Chen, S. C., Ma, Z. T., 2006. High order generalized screen propagator for wave equation prestack depth migration. Chinese Journal of Geophysics 49 (5), 1290–1297.

Claerbout, J., 1985. Imaging the Earth's interior. Blackwell Scientific Publications, Inc.

Collino, F., Joly, P., 1995. Splitting of operators, alternate directions and paraxial approximations for the three-dimensional wave equation. SIAM Journal on Scientific Computing 16, 1019–1048.

Gazdag, J., Sguazzero, P., 1984. Migration of seismic data by phase shift plus interpolation. Geophysics 49 (2), 124–131.

Gray, S. H., Etgen, J., Dellinger, J., Whitmore, D., 2001. Y2K Review Article: Seismic migration problems and solutions. Geophysics 66, 1622–1640.

Hale, D., 1991. 3-D depth migration via McClellan transforms. Geophysics 56, 1778–1785.

Han, B., Gu, H., Liu, S., Yan, Z., Tang, Y., Liu, C., 2018. Wavelength-dependent Fresnel beam propagator and migration in VTI media. Journal of Applied Geophysics 155, 176–186.

Hua, B.L., Calandra, H., Williamson, P., 2006. 3D common azimuth Fourier finite-difference depth migration in transversely isotropic media. 76th Annual International Meeting, SEG, Expanded Abstracts, 2387–2391.

Huang, L.-J., Fehler, M. C., Wu, R.-S., 1999. Extended local Born Fourier migration method. Geophysics 64 (5), 1524–1534.

Jang, S., Kim, T., 2016. Prestack Depth migration by a Parallel 3D PSPI: International Journal of Geosciences 7, 904–914.

Jin, S, Wu, R. S., 1999. Common-offset pseudo-screen depth migration. 69th Annual International Meeting, SEG, Expanded Abstracts, 1516 –1519.

Le Rousseau, J.H., de Hoop, M. V., 2001a. Modeling and imaging with the scalar generalized-screen algorithms in isotropic media. Geophysics 66, 1551–1568.

Le Rousseau, J. H., de Hoop, M. V., 2001b. Scalar generalized-screen algorithms in transversely isotropic media with a vertical symmetry axis. Geophysics 66, 1538–1550.

Li, J., Fehler, M., Yang, D., Huang, X., 2015. 3D weak-dispersion reverse time migration using a stereo-modeling operator. Geophysics 80, S19–S30.

Liu, B., Sacchi, M.D., 2004. Minimum weighted norm interpolation of seismic records. Geophysics 69, 1560-1568.

Margrave, G. F., Ferguson, R. J., 1999. Wavefield extrapolation by nonstationary phase shift. Geophysics 64 (4), 1067–1078.

Margrave, G. F., 1998. Theory of nonstationary linear filtering in the Fourier domain with application to time-variant filtering. Geophysics 63 (1), 244–259.

Popovici, A.M., 1996. Prestack migration by split-step DSR. Geophysics 61, 1412–1416.

Ristow, D., Rühl, T., 1994. Fourier Finite difference migration. Geophysics 59 (12), 1882–1893.

Shan, G., 2009. Optimized implicit finite-difference and Fourier finite-difference migration for VTI media. Geophysics 74, WCA189-WCA197.

Shin, S., Byun, J., 2013. Development of a Prestack Generalized-Screen Migration Module for Vertical Transversely Isotropic Media. Jigu-Mulli-wa-Mulli-Tamsa 16, 71–78.

Stoffa, P. L., Fokkerna, J. T., Luna Freire, R. M., Kessinger, W. P., 1990. Split-step Fourier migration. Geophysics 55 (4), 410–421.

Xu, S., Zhang, Y., Pham, D., Lambare, G., 2005. Antileakage Fourier transform for seismic data regularization. Geophysics 70, V87-V95.

Zhang, Y., Notfors, C., Xie, Y., 2003. Stable xk wavefield extrapolation in a v(x,y,z) medium. SEG 74th Annual Meeting, Expanded Abstracts.

Zhang, J.-H., Yao, Z.-X., 2012. Globally optimized finite-difference extrapolator for strongly VTI media. Geophysics 77, T125–T135.

# LIST OF FIGURES

LIST OF TABLES